\documentclass[graybox]{svmult}

\usepackage{mathptmx}       
\usepackage{helvet}         
\usepackage{courier}        
\usepackage{type1cm}        
\usepackage{makeidx}         
\usepackage{graphicx}        
\usepackage{multicol}        
\usepackage[bottom]{footmisc}

\makeindex             

\newcommand\rf[1]{(\ref{eq:#1})}
\newcommand\lab[1]{\label{eq:#1}}

\newcommand\br{\begin{eqnarray}}
\newcommand\er{\end{eqnarray}}
\newcommand\be{\begin{equation}}
\newcommand\ee{\end{equation}}

\newcommand\lb{\lbrack}
\newcommand\rb{\rbrack}

\renewcommand\){\right)}

\newcommand\bc{\begin{center}}
\newcommand\ec{\end{center}}

\newcommand\partder[2]{\frac{{\partial {#1}}}{{\partial {#2}}}}

\renewcommand\d{\delta}
\newcommand\eps{\epsilon}
\newcommand\vareps{\varepsilon}

\newcommand\h{\frac{1}{2}}
\renewcommand\k{\kappa}
\renewcommand\l{\lambda}
\renewcommand\L{\Lambda}
\newcommand\m{\mu}
\newcommand\n{\nu}
\newcommand\om{\omega}

\newcommand\vp{\varphi}
\renewcommand\P{\Phi}
\newcommand\pa{\partial}

\renewcommand\r{\rho}

\renewcommand\th{\theta}

\newcommand\wti{\widetilde}

\newcommand\cB{{\mathcal B}}

\newcommand\cE{{\mathcal E}}

\newcommand\cH{{\mathcal H}}

\newcommand{\ct}[1]{\cite{#1}}

\newcommand\PRL[3]{\textsl{Phys. Rev. Lett.} \textbf{#1}, #3 (#2)}

\newcommand\PRD[3]{\textsl{Phys. Rev.} \textbf{D#1}, #3 (#2)}

\newcommand\PLB[3]{\textsl{Phys. Lett.} \textbf{#1B}, #3 (#2)}

\newcommand\IJMPA[3]{\textsl{Int. J. Mod. Phys.} \textbf{A#1}, #3 (#2)}

\newcommand\vpdot{\stackrel{.}{\varphi}}

\newcommand\adot{\stackrel{.}{a}}

\begin{document}

\title*{Cosmology via Metric-Independent Volume-Form Dynamics}
\author{Eduardo Guendelman and Emil Nissimov and Svetlana Pacheva}
\institute{Eduardo Guendelman \at Department of Physics, Ben-Gurion University of
the Negev, Beer-Sheva, Israel \email{guendel@bgu.ac.il}
\and Emil Nissimov \at Institute for Nuclear Research and Nuclear Energy,
Bulgarian Academy of Sciences, Sofia, Bulgaria \email{nissimov@inrne.bas.bg}
\and Svetlana Pacheva \at Institute for Nuclear Research and Nuclear Energy,
Bulgarian Academy of Sciences, Sofia, Bulgaria \email{svetlana@inrne.bas.bg}}
%
%
\maketitle

\abstract*{The method of non-Riemannian volume-forms (metric-independent covariant 
integration measure densities on the spacetime manifold) is applied to
construct a unified model of dynamical dark energy plus dark matter as a
dust fluid flowing along geodesics, which results from a hidden Noether 
symmetry of the pertinent scalar field Lagrangian. Canonical Hamiltonian treatment 
and Wheeler-DeWitt quantization of the latter model are briefly discussed.}

\abstract{The method of non-Riemannian volume-forms (metric-independent covariant 
integration measure densities on the spacetime manifold) is applied to
construct a unified model of dynamical dark energy plus dark matter as a
dust fluid resulting from a hidden Noether symmetry of the pertinent scalar
field Lagrangian. Canonical Hamiltonian treatment and Wheeler-DeWitt
quantization of the latter model are briefly discussed.}

\section{Introduction}
\label{intro}

Alternative spacetime volume-forms (generally-covariant
integration measure densities) independent on the Riemannian
metric on the pertinent spacetime manifold have profound impact in any
field theory models with general coordinate reparametrization invariance, 
such as general relativity and its extensions, strings and (higher-dimensional) 
membranes \ct{TMT-orig-1,TMT-orig-2,TMT-orig-3,mstring}. 

The principal idea is to replace or employ alongside the standard Riemannian
integration density given by $\sqrt{-g}$ (square root of the determinant 
$g=\det\Vert g_{\m\n}\Vert$ of the Riemannian metric $g_{\m\n}$) 
one or more non-Riemannian
(metric-independent) covariant integration measure densities defined in
terms of dual field-strengths $\P(B)$ of auxiliary maximal rank antisymmetric 
tensor gauge fields $B_{\m\n\l}$:
\be
\P(B) = \frac{1}{3!}\vareps^{\m\n\k\l} \pa_\m B_{\n\k\l} \; ,
\lab{mod-measure}
\ee
The corresponding non-Riemannian-modified-measure 
gravity-matter models were called ``two-measure (gravity) theories'' and the 
associated auxiliary tensor gauge fields $B_{\m\n\l}$ -- ``measure gauge fields''.

The auxiliary ``measure'' gauge fields trigger a number of physically interesting
phenomena:

\begin{itemize}
\item
The equations of motion w.r.t. $B_{\m\n\l}$
produce dynamical constraints involving  {\em arbitrary integration constants},
where one of the latter {\em always} acquires the meaning of a 
{\em dynamically generated cosmological constant}.
\item
Employing the canonical Hamiltonian formalism for Dirac-constrained
systems we find that $B_{\m\n\l}$ 
are in fact
almost pure gauge degrees of freedom except for the above mentioned
arbitrary integration constants which are identified with the conserved
Dirac-constrained canonical momenta conjugated to the ``magnetic'' components 
$(B_{ijk})$ of the ``measure'' gauge fields. 
\item
Upon applying the non-Riemannian volume-form formalism to minimal $N=1$ 
supergravity the dynamically generated cosmological constant triggers spontaneous
supersymmetry breaking and mass generation for the gravitino (supersymmetric 
Brout-Englert-Higgs effect) \ct{susyssb}. Applying the same 
formalism to anti-de Sitter supergravity allows to produce
simultaneously a very large physical gravitino mass and a very small 
{\em positive} observable cosmological constant \ct{susyssb} in accordance 
with modern cosmological scenarios for slowly expanding universe of the present epoch
\ct{dark-energy-observ-1,dark-energy-observ-2,dark-energy-observ-3}. 
\item
Employing two independent non-Riemannian volume-forms like \rf{mod-measure} 
in generalized gravity-gauge+scalar-field models \ct{emergent}, 
thanks to the appearance of several arbitrary integration constants through the 
equations of motion w.r.t. the ``measure'' gauge fields, we obtain in the
physical ``Einstein-frame'' a
remarkable effective scalar potential with two infinitely large flat regions
(for large negative and large positive values of the scalar field $\vp$) with vastly
different scales appropriate for a unified description of both the early and
late universe' evolution.
Another remarkable feature is the existence of a stable initial phase of
{\em non-singular} universe creation preceding the inflationary phase
-- stable ``emergent universe'' without ``Big-Bang'' \ct{emergent}.
\end{itemize}

As a specific illustration of the usefulness of the non-Riemannian volume-form method
and extending the study in \ct{eduardo-singleton,eduardo-ansoldi} we discuss
a modified gravity+single-scalar-field model where the scalar
Lagrangian couples symmetrically both to the standard Riemannian volume-form
given by $\sqrt{-g}$ as well as to another non-Riemannian volume-form \rf{mod-measure}.
The pertinent scalar field dynamics provides a unified description of both dark energy
via dynamical generation of a cosmological constant, and dark matter as a
``dust'' fluid with geodesic flow as a result of a hidden Noether symmetry.
Further, we briefly consider the canonical Hamiltonian treatment and the
Wheeler-DeWitt quantization of the above unified dark energy plus dust fluid dark
matter model.

\section{Dark Energy and Dust Fluid Dark Matter via Non-Riemannian Volume-Form 
Dynamics} 
\label{TMT}

We will consider the following non-conventional gravity+scalar-field
action -- a particular case of the general class of the ``two-measure''
gravity-matter theories \ct{TMT-orig-1,TMT-orig-2,TMT-orig-3}
(for simplicity we use units with the Newton constant $G_N = 1/16\pi$):
\be
S = \int d^4 x \sqrt{-g}\, R +
\int d^4 x \bigl(\sqrt{-g}+\P(B)\bigr) L(\vp,X) \; .
\lab{TMT}
\ee
Here $\P (B)$ is as in \rf{mod-measure} and $L(\vp,X)$ is general-coordinate 
invariant Lagrangian of a single scalar field $\vp (x)$ of a generic ``k-essence'' 
form \ct{k-essence-1,k-essence-2} (i.e., a nonlinear (in general) function of the scalar
kinetic term $X$): 
$L(\vp,X) = \sum_{n=1}^N A_n (\vp) X^n - V(\vp) \;\; ,\;\;
X \equiv - \h g^{\m\n}\pa_\m \vp \pa_\n \vp$.
The energy-monentum tensor corresponding to \rf{TMT} reads:
\be
T_{\m\n} = g_{\m\n}\, L(\vp,X) + 
\Bigl( 1+\frac{\P(B)}{\sqrt{-g}}\Bigr) \partder{L}{X}\, \pa_\m \vp\, \pa_\n \vp \; .
\lab{EM-tensor} 
\ee
The essential new feature is the dynamical constraint on the scalar
Lagrangian, which results from the equation of motion w.r.t. ``measure'' 
gauge field $B_{\m\n\l}$:
\be
\pa_\m L (\vp,X) = 0 \;\;\; \longrightarrow \;\;\;
L (\vp,X) = - 2M = {\rm const} \; ,
\lab{L-const}
\ee
where $M$ is an {\em arbitrary integration constant}. We will take $M>0$ in 
view of its interpretation as a {\em dynamically generated cosmological constant}
(see \rf{T-J-hydro} below).

A remarkable property of the scalar field action in \rf{TMT} is the presence
of a hidden Noether symmetry of the latter under the nonlinear transformations:
\be
\d_\eps \vp = \eps \sqrt{X} \;\; ,\;\;\d_\eps g_{\m\n} = 0 \;\; ,\;\;
\d_\eps B_{\m\n\l} = - \eps \frac{1}{2\sqrt{X}} \vareps_{\m\n\l\k}
g^{\k\r}\pa_\r\, \vp \bigl(\P(B) + \sqrt{-g}\bigr)  \; .
\lab{hidden-sym}
\ee
The standard Noether procedure yields the conserved current:
\be
\nabla_\m J^\m = 0 \quad ,\quad
J^\m \equiv \Bigl(1+\frac{\P(B)}{\sqrt{-g}}\Bigr)\sqrt{2X} 
g^{\m\n}\pa_\n \vp \partder{L}{X} \; .
\lab{J-conserv}
\ee
Let us stress that the existence of the hidden symmetry
\rf{hidden-sym} of the action \rf{TMT} {\em does not} depend on the specific
form of the scalar field Lagrangian. 

Now, $T_{\m\n}$ \rf{EM-tensor} and $J^\m$ \rf{J-conserv} can be rewritten in
a relativistic hydrodynamical form (taking into account \rf{L-const}):
\be
T_{\m\n} = \rho_0 u_\m u_\n - 2M g_{\m\n} \quad ,\quad J^\m = \rho_0 u^\m \; ,
\lab{T-J-hydro}
\ee
where:
\be
\rho_0 \equiv \Bigl(1+\frac{\P(B)}{\sqrt{-g}}\Bigr)\, 2X \partder{L}{X} 
\quad,\quad
u_\m \equiv \frac{\pa_\m \vp}{\sqrt{2X}} \quad (\; u^\m u_\m = -1\;) \; .
\lab{rho-u-def}
\ee
For the pressure $p$ and energy density $\rho$ we obtain:
\be
p = - 2M = {\rm const} \quad ,\quad
\rho = \rho_0 - p = 2M + \Bigl(1+\frac{\P(B)}{\sqrt{-g}}\Bigr)\, 2X \partder{L}{X}
\; ,
\lab{p-const-rho-def}
\ee
wherefrom indeed the integration constant $M$ 
appears as {\em dynamically generated cosmological constant}.
Moreover the covariant energy-momentum conservation $\nabla^\n T_{\m\n}=0$,
due to the constancy of the pressure (first Eq.\rf{p-const-rho-def}),
actually implies {\em both} the conservation of the Noether current $J^\m$
\rf{J-conserv} as well as the {\em geodesic flow} equation:
$~u_\n \nabla^\n u_\m = 0 \;$ .

The above results lead to the following interpretation in accordance with the  
standard $\L$-CDM model (see e.g. \ct{Lambda-CDM}). 
The energy-momentum tensor \rf{T-J-hydro} consists of two parts:

\begin{itemize}
\item
Dark energy part given by the second cosmological constant term in $T_{\m\n}$
\rf{T-J-hydro}, which arises due to the dynamical constraint on the scalar field
Lagrangian \rf{L-const} 
with $p_{\rm DE} = -2M\, ,\, \rho_{\rm DE} = 2M$ (cf. Eqs.\rf{p-const-rho-def}).
\item
Dark matter part given by the first term in \rf{T-J-hydro}
(cf. also \rf{p-const-rho-def}) with $p_{\rm DM} = 0\, ,\, \rho_{\rm DM} = \rho_0$
($\rho_0$ as in \rf{rho-u-def}). The latter describe a dust fluid with dust 
``particle number'' conservation \rf{J-conserv} and flowing along geodesics.
\end{itemize}

The idea of unified description of dark energy and dark matter is the subject
of numerous earlier papers exploiting a variety of different approaches. Among
them are generalized Chaplygin gas models \ct{chaplygin-1,chaplygin-2}, ``mimetic'' dark
matter models \ct{mimetic-1,mimetic-2}, 
constant pressure ansatz models \ct{cruz-etal} \textsl{etc.} 

\section{Canonical Hamiltonian Formalism and Wheller-DeWitt Equation}
\label{WDW}

For a systematic canonical Hamiltonian treatment of gravity-matter models
based on metric-independent volume-forms we refer to \ct{grav-bags-dusty} and
specifically to the second reference therein for the full Hamiltonian treatment 
of the present
model \rf{TMT}. Here, for simplicity, we
will consider a reduction of \rf{TMT} where the spacetime metric
is taken of the Friedmann-Lemaitre-Robinson-Walker (FLRW) class:
 \be
ds^2 = - N^2(t) dt^2 + a^2(t) \Bigl\lb \frac{dr^2}{1-K r^2}
+ r^2 (d\th^2 + \sin^2\th d\phi^2)\Bigr\rb \; ,
\lab{FLRW}
\ee
and where $\vp$ and the ``measure'' gauge field $B$ are taken to depend only on $t$.
The reduced action resulting from \rf{TMT} reads (taking the standard form
of the scalar Lagrangian): 
\be
S = 6 \int dt N a^3 \Bigl\lb - \frac{1}{N^2}\Bigl(\frac{\adot}{a}\Bigr)^2 + 
\frac{K}{a^2}\Bigr\rb 
+ \int dt \bigl( \pa_t B + N a^3\bigr) 
\bigl(\frac{1}{2N^2}\vpdot^2 - V(\vp)\bigr) \; .
\lab{TMT-reduced}
\ee
The equation of motion w.r.t. $B$ produces the dynamical constraint (reduced
form of \rf{L-const}) with explicit solution for $\vp (t)$:
\be
\vpdot^2 = 2\bigl( V(\vp) - 2M\bigr) \;\; \longrightarrow \;\;
\int_{\vp (0)}^{\vp (t)} \frac{d\vp}{\sqrt{2\bigl( V(\vp) - 2M\bigr)}} = \pm t .
\lab{vp-eq-2-0} 
\ee
The hidden ``dust'' Noether symmetry (cf. \rf{hidden-sym} and \rf{J-conserv}) 
of the reduced action \rf{TMT-reduced} now takes the form:
\be
\d_\eps \vp = \eps \frac{\vpdot}{N} \;\;,\;\;
\d_\eps B = \eps \frac{1}{N} \bigl(\pa_t B + N a^3\bigr) \;\; ,\;\; \d_\eps a =0 
\quad ,\;\;\;
\frac{d}{dt}\Bigl\lb \(N a^3 + \pa_t B\) \frac{\vpdot^2}{N^3}\Bigr\rb = 0 \; .
\lab{J-conserv-reduced}
\ee
The canonical Hamiltonian treatment a'la Dirac of the reduced action
\rf{TMT-reduced} yields the following Dirac-constrained Hamiltonian
($N$ appearing as a Lagrange multiplier of the first class constraint in the
brackets):
\be
\cH_{\rm total} = N \Bigl\lb - \frac{p_a^2}{24a} - 6Ka - \pi_B a^3 +
\sqrt{2\bigl( V(\vp) + \pi_B\bigr)}\, p_\vp \Bigr\rb \; ,
\lab{can-Ham-reduced}
\ee
where $p_a$ and $\pi_B$ are the canonically conjugated momenta of $a$ and
$B$, respectively.

The quantum Wheeler-DeWitt equation corresponding to \rf{can-Ham-reduced} is
significantly simplified upon changing variables as:
\be
a \to {\wti a} = \frac{4}{\sqrt{3}}\, a^{3/2} \quad ,\quad
\vp \to {\wti \vp} = \int \frac{d\vp}{\sqrt{2\bigl( V(\vp)-2M\bigr)}} \; ,
\lab{new-var}
\ee
where from \rf{vp-eq-2-0} we find that the new scalar field coordinate
${\wti \vp}$ will have the meaning of a (cosmic) time. 
Since $B$ turns out to be a cyclic variable in \rf{can-Ham-reduced} the
quantized canonical momentum ${\widehat \pi}_B = - i \d/\d B$ is
immediately diagonalized whose eigenvalues are denoted by $\pi_\cB = - 2M$,
so that $M$ will have the meaning of a dynamically generated cosmological
constant. Further, we notice
that the quantized form of the last term in \rf{can-Ham-reduced}, which is
the Hamiltonian expression for the conserved ``dust'' Noether symmetry charge
\rf{J-conserv-reduced}, will simplify to
$\sqrt{2\bigl( V(\vp) + \pi_B\bigr)}\, \bigl(-i \frac{d}{d \vp}\bigr) =
-i d/d {\wti \vp}$
and is straightforwardly diagonalized with eigenvalues $\cE$. 
Accordingly, the total Wheeler-DeWitt wave function will have the
form 
$\psi (a,\vp,B) = \psi_{\rm grav} ({\wti a})\, e^{i\cE {\wti \vp}- i\,2M B}$
(with ${\wti a}$ and ${\wti \vp}$ as in \rf{new-var}),
and the Wheeler-DeWitt equation reduces to ``energy'' eigenvalue Schr{\"o}dinger 
equation for the gravitational part of the total wave function: 
\be
\Bigl\lb -\h \frac{\pa^2}{\pa {\wti a}^2} - \frac{3}{8}M {\wti a}^2
+6K \bigl(\frac{\sqrt{3}}{4}\,{\wti a}\bigr)^{2/3} - \cE \Bigr\rb 
\psi_{\rm grav} ({\wti a}) = 0 
\lab{WDW-reduced-E}
\ee
In the special case of zero spacial curvature $K=0$ in the FLRW metric \rf{FLRW}
Eq.\rf{WDW-reduced-E} reduces to the energy eigenvalue Schr{\"o}dinger equation 
for the {\em inverted} harmonic oscillator \ct{inverted-osc}
with negative frequency squared $\om^2 \equiv - \frac{3}{4}\, M$
(the dynamically generated cosmological constant $M$ must be positive). 

In particular, the inverted oscillator was applied in \ct{guth-pi}
to study the quantum mechanical dynamics of the scalar field in the so
called ``new inflationary'' scenario. Since the energy eigenvalue spectrum 
of the inverted harmonic oscillator is continuous ($\cE \in (-\infty,+\infty)$)
and the corresponding energy eigenfunctions are not square-integrable, its
application in the context of cosmology \ct{guth-pi} requires employment of
wave-packets instead of energy eigenfunctions.

\begin{acknowledgement}
E.G. thanks Frankfurt Institute for Advanced Studies (FIAS) for hospitality.
We gratefully acknowledge support of our collaboration through the 
academic exchange agreement between the Ben-Gurion University in Beer-Sheva,
Israel, and the Bulgarian Academy of Sciences. 
S.P. and E.N. have received partial support from European COST actions
MP-1210 and MP-1405, respectively, as well from Bulgarian National Science
Fund Grant DFNI-T02/6. 
\end{acknowledgement}



\end{document}